\documentclass[12pt]{article}
\usepackage{amssymb}
\usepackage{graphicx}%
\usepackage{amsfonts}%
\usepackage{amssymb}
\usepackage{psfrag}
\newcommand{\bea}{\begin{eqnarray}}
\newcommand{\eea}{\end{eqnarray}}
\newcommand{\be}{\begin{equation}}
\newcommand{\ee}{\end{equation}}
\newcommand{\vs}[1]{\vspace{#1 mm}}

\newcommand{\dsl}{\pa \kern-0.5em /}

\newcommand{\pa}{\partial}

\newcommand{\nn}{\nonumber\\}

\begin{document}
\topmargin 0pt
\oddsidemargin 0mm



\begin{flushright}

USTC-ICTS-08-17\\

\end{flushright}

\vspace{2mm}

\begin{center}

{\Large \bf On the low energy brane/anti-brane dynamics

}

\vs{10}

{\large J. X. Lu\footnote{E-mail: jxlu@ustc.edu.cn}, Bo
Ning\footnote{E-mail: nwaves@mail.ustc.edu.cn}
 and Guan-Nan Zhong\footnote{E-mail: zhonggn@email.jlu.edu.cn}}

 \vspace{6mm}

{\em

  Interdisciplinary Center for Theoretical Study\\

 University of Science and Technology of China, Hefei, Anhui
 230026, China\\

}

\end{center}

\vs{9}

\begin{abstract}

We study the dynamical behavior of a pair of Dp-brane and anti
Dp-brane ($0 \leq p \leq 6$) moving parallel to each other in the
region where the brane and anti-brane annihilation will not occur
and the low energy description is valid. Given this, we perform a
general analysis, in the center of mass frame, of the behavior of
the effective potential with respect to the relative brane
separation and find that the classical orbits of this system are in
general unbound except for $p = 6$ case for which classical  bound
orbits exist. The non-linearity of the low energy DBI action for
D-brane is important for the underlying dynamics. We solve also the
explicit orbits for $p = 6$ case.

\end{abstract}
\newpage

\vs{15}

\section{Introduction}

It is known that two parallel D-branes separated by a distance feels
no force between them, independent of their separation. This is due
to the BPS nature or the preservation of certain number of
space-time supersymmetries of this system, and goes by the name
``no-force" condition. This was shown initially for brane
supergravity configurations through a probe
\cite{Dabholkar:1990yf,Duff:1994an} and later through the full
string level computations as an open string one-loop annulus diagram
with one end of the string located at one D-brane and the other end
at the other D-brane making use of the ``usual abstruse identity"
\cite{Polchinski:1995mt}. With this, one can easily infer that when
one of branes in the above is replaced by the corresponding
anti-brane, there must be a separation-dependent non-vanishing force
to arise since such a system breaks all the space-time
supersymmetry. The corresponding forces can easily be computed given
our knowledge of computing forces between two identical branes.

Analyzing the force behavior on the brane separation indicates the
well-known fact that the brane and anti-brane will start to
annihilate each other when the separation is on the order of string
scale\cite{Banks:95fc, Lu:2008fc}. Therefore to prevent this from
happening, we should limit the brane separation to be much larger
than the string length $l_s = \sqrt{\alpha'}$ with the string
parameter $\alpha'$ related to the tension $T$  by $\alpha' =
1/(2\pi T)$. When this is met, the attractive force between the
brane and the anti-brane can well be approximated by its
long-distance form and the brane or anti brane itself is believed to
be described by its low energy Dirac-Born-Infeld (DBI) action. So we
have a well-defined system of D-brane and anti D-brane plus their
interaction in the region where the brane and anti-brane
annihilation will not occur and the low energy description is valid.

The DBI action, describing  the low energy behavior of D-brane, is
nonlinear and it is interesting to explore how D-brane behaves under
this action. For this purpose, we analyze the dynamical behavior of
a pair of Dp-brane and anti Dp-brane ($0 \leq p \leq 6$) placed
parallel to each other under the conditions mentioned
above\footnote{A probe approach with similar considerations has been
given in \cite{Burgess:2003qv, Burgess:2003mm} where the orbits of a
probe anti $p'$-brane in a background of a stack of $N$ Dp branes
with either $p' = p$ or $p' \neq p$ and $g_s N \gg 1$ have been
considered, respectively. Our focus here is on the classical orbits
of a pair of Dp-brane and anti Dp-brane in a different region of
$g_s N \ll 1$ with $N = 1$ and a flat background.  Even so, many
conclusions drawn here such as the $p = 6$ case is singled out for
the existence of bound orbits are the same as theirs.}. In section
2, we set up the system so we can perform an analysis in the later
sections. In section 3, we give a general analysis based on the
effective potential and find that the classical orbits are in
general unbound except for $p = 6$ case where classical bound orbits
exist. We also derive the conditions for which the respective orbits
can exist. In section 4, we solve the explicit orbits for the $p =
6$ case as an example and find that they are consistent with our
general analysis given in the previous section. We conclude this
paper in section 5.

\section{Preliminaries}

We consider a system consisting of a pair of Dp-brane and anti
Dp-brane ($0 \leq p \leq 6$) moving parallel to each other in the
region where the brane and anti-brane annihilation will not occur
and the low energy description for the branes is valid. Given this,
even though the brane tension is now large due to the weak string
coupling limit, the spacetime around each brane still remains as
flat so long we don't probe a distance much smaller than the string
scale \cite{Lu:2008fc}. Our interest is in the classical relative
motion of the two branes under their mutual attractive long-distance
interaction and so we don't expect to excite the gauge modes on the
brane. So the DBI action for either of the branes is reduced to

\be \label{dbiact}\textsc{S}_{\rm DBI} = -T_p \int d^{p + 1} \sigma
\sqrt{- \det \gamma_{\mu\nu}},\ee where the induced metric on the
brane is  $\gamma_{\mu\nu} = \partial_\mu X^M
\partial_\nu X^N \eta_{MN} $ with $\eta_{MN}$ the spacetime Minkowski
flat metric and the brane tension $T_p= 1 /[g_s
(2\pi)^p\,\alpha\,'^{\frac{p+1}{2}}]$ with $g_s$ the string
coupling. We take now the so-called static gauge, i.e. $\sigma^\mu =
X^\mu$ with $\mu = 0, 1, \cdots p$ and the underlying dynamics under
consideration says that the coordinates transverse to the brane
depend only on time, i.e., $X^i = X^i (\tau)$ with $\tau = \sigma^0
= t$ and $ i=p+1,...,9$. If we denote $\dot{X}^i = d X^i/d\tau$, we
have from the above the following \be\label{rdbiact}  S_{\rm DBI} =
- T_p \,V_p \,\int d\tau \,\sqrt{1-\dot{X^i}\dot{X^i}}, \ee where
the brane spatial volume $V_p = \int d^p \sigma$. In the following,
for simplicity, we also use the vector notation $\dot{\vec{X}}$ in
replace of its components $\dot{X}^i$.

At any moment $\tau$, we denote the Dp-brane and the anti Dp-brane
locations in the direction transverse to the branes as $\vec{X}_1$
and $\vec{X}_2$, respectively. With this, the interaction potential
between them at a separation $r = |\vec{X}_2 - \vec{X}_1|$ can be
obtained the way described in the Introduction and is given by

\be
 V = -V_{p + 1} \int_0^\infty\frac{ds}{s} \left(2\pi s\right)^{- \frac{ p + 1}{2}} e^{-
\frac{s}{4\alpha'}( \frac{r^2}{2\pi^2\alpha'} - 1) } \prod_{n =
1}^\infty \left(\frac{1 - q^{2n - 1}}{1 - q^{2 n}}\right)^8,
\label{ptl}\ee  where the $\rm p$-brane worldvolume  $V_{p + 1} =
V_p \int d\tau$, $q = e^{- s/4\alpha'}$ and the integration variable
$s$ is the proper time in the open string channel. When the
separation $r$ approaches the string scale, the interaction force
will be divergent, indicating the occurring of the tachyon
condensation \cite{Banks:95fc, Lu:2008fc}. For the interest of this
paper, we have $r \gg l_s = \sqrt{\alpha'}$. So the main
contribution to the integral in the above potential comes from small
$s$. Note that \cite{Lu:2008fc}, \be g (s) = \prod_{n = 1}^\infty
\left(\frac{1 - q^{2n - 1}}{1 - q^{2 n}}\right)^8 \rightarrow (
s/2\pi\alpha\,')^4 \qquad {\rm as} \qquad s\to 0,\ee So the
long-distance potential is attractive, as expected, and is given as
$ V (r) = - V_{p+1}~ \frac{\Omega}{r^{7-p}}$ with $\Omega ~=
~2\,^{2(3-p)}~\pi^{\frac{5-3p}{2}}~\alpha\,'\,^{3-p}~\Gamma
\left(\,\frac{ 7-p}{2}\,\right)$. Therefore the complete action for
the Dp-$\bar{\rm D}$p system ($0 \leq p \leq 6$) in the region of
interest is: \be \textsc{S} = - T_p\,V_p\int d\tau\left(
\sqrt{1-\dot{\vec{X}_1}\cdot\dot{\vec{X}_1}}+\sqrt{1-\dot{\vec{X}_2}\cdot\dot{\vec{X}_2}}~\right)
 ~+~ V_p\int d\tau \frac{\Omega}{~~|\vec{X_2}-\vec{X_1}|^{7-p}} \,,
\label{act1}\ee  which looks like a classical two-body mechanical
problem with now the non-linear kinetic terms. To simplify the
discussion, as usual for a two-body system, we choose the center of
mass coordinate $\vec{R}$ and the relative coordinate $\vec{r} = 2\,
\vec{\rho}$ as \bea \vec{R}&=& \frac{\vec{X}_1+\vec{X}_2}{2},
\qquad\qquad\qquad \vec{X}_1=\vec{R}+\frac{\vec{r}}{2} = \vec{R} +
\vec{\rho}\nn \vec{r}= 2 \vec{\rho} &=& \vec{X}_1-\vec{X}_2, \qquad
\qquad\qquad \vec{X}_2=\vec{R}-\frac{\vec{r}}{2} = \vec{R} -
\vec{\rho}\,\,\,,\eea where for later convenience, we have also
introduced  $\vec{\rho}$ which is half of the relative separation
vector $\vec{r}$. Then the above action, denoting
$\rho=|\,\vec{\rho}\,|$, becomes

\bea \textsc{S}   &=& - T_p\,V_p\int d\tau\left[\,
\sqrt{1-(\dot{\vec{R}}+\dot{\vec{\rho}}\,)\cdot(\dot{\vec{R}}+\dot{\vec{\rho}}\,)}\,\,+
\,\sqrt{1-(\dot{\vec{R}}-\dot{\vec{\rho}}\,)\cdot(\dot{\vec{R}}-\dot{\vec{\rho}}\,)}~\right]\nn
&& ~+~ V_p\int d\tau \frac{\Omega}{~(2\rho)^{7-p}}.\label{act2}\eea
 Note that the generalized momentum corresponding to coordinate
$\vec{R}$\, is conserved:

\be \vec{P}_{\vec{R}}={\partial \,\textsc{L }\over\partial
\dot{\vec{R}}}=V_p\, T_p
\left(\frac{\dot{\vec{R}}+\dot{\vec{\rho}}}{\sqrt{1-(\dot{\vec{R}}+\dot{\vec{\rho}}\,)\cdot(\dot{\vec{R}}+\dot{\vec{\rho}}\,)}}+
\frac{\dot{\vec{R}}-\dot{\vec{\rho}}}{\sqrt{1-(\dot{\vec{R}}-\dot{\vec{\rho}}\,)\cdot(\dot{\vec{R}}-\dot{\vec{\rho}}\,)}}
\right)\equiv \vec{c}\,\,,\label{cons} \ee  where $\vec{c}$ ~is a
constant vector. The equation of motion with respect to $\vec{\rho}$
is

\be T_p ~\frac{d}{d\tau}~\left(
\frac{\dot{\vec{R}}+\dot{\vec{\rho}}}{\sqrt{1-(\dot{\vec{R}}+\dot{\vec{\rho}}\,)
\cdot(\dot{\vec{R}}+\dot{\vec{\rho}}\,)}}\, -
\frac{\dot{\vec{R}}-\dot{\vec{\rho}}}{\sqrt{1-(\dot{\vec{R}}-\dot{\vec{\rho}}\,)
\cdot(\dot{\vec{R}}-\dot{\vec{\rho}}\,)}}\right)~ + 2
\frac{\Lambda}{\rho^{9-p}} ~~ \vec{\rho} = 0\label{eom1} \ee with $
\label{biglambda} \Lambda ~=~
\frac{1}{2^{7-p}}\,\frac{7-p}{2}\,\Omega~=
~2\,^{-(1+p\,)}~\pi^{\frac{5-3p}{2}}~\alpha\,'\,^{3-p}~\Gamma
\left(\,\frac{ 9-p}{2}\,\right)$.

In what follows, we focus on the case $\dot{\vec{R}} = 0$ at a given
initial time $\tau_i$, i.e. in the center of mass frame, implying
$\vec{c} = 0$ in (\ref{cons}).   One can then show from the same
equation $\dot{\vec{R}} = 0$ at all time. This is entirely
consistent with our usual picture that if we are in the center of
mass frame at a given time, we are always so if only mutual
interactions are present\footnote{If we are not in the center of
mass frame, the underlying dynamics seems rather complicated due to
the non-linear DBI action and we try to explore this case
elsewhere.}.

For $\dot{\vec{R}}=0$, all the  equations of motion reduce to the
following one \be T_p ~\frac{d}{dt}~\left(
\frac{\dot{\vec{\rho}}}{\sqrt{1-\dot{\vec{\rho}}~^2}}\right)~
+\,\frac{\Lambda}{\rho^{9-p}}~ \vec{\rho} ~=~ 0.\label{EOM}\ee This
can be easily integrated out to give
 \be \frac{1}{\sqrt{1-\dot{\vec{\rho}}~^2}}~ -
\frac{\lambda}{\rho^{7-p}}~
 =~ H ~,\label{engy1}\ee  where $H$ is an integral constant and
 $ \label{littlelambda} \lambda ~=~
 \Lambda /[T_p (7-p)] ~=~ \frac{g_s}{4\pi}~(\pi \alpha\,'\,)^
 {\frac{7-p}{2}}~\Gamma \left(\frac{7-p}{2}\right)\sim g_s \,\alpha'^{\frac{7 -
 p}{2}}$.
The integral constant $H$ is actually the reduced Hamiltonian in the
sense defined below: the Lagrangian from (\ref{act2}) when
$\dot{\vec{R}} = 0$ is \be \label{lag} \textsc{L} = -  2 T_p V_p
\sqrt{1 - {\dot{\vec{\rho}}}^2} + \frac{V_p}{2^{7 - p}}
\frac{\Omega}{\rho^{7 - p}},\ee and then we have the Hamiltonian \be
\textsc{H} \, =\, \vec{P}_{\vec{\rho}}\cdot \dot{\vec{\rho}} \,-\,
\textsc{L} \,=\, 2 \,V_p \,T_p \left(
\frac{1}{\sqrt{1-\dot{\vec{\rho}}~^2}}~ - \frac{\lambda}{\rho^{7-p}}
\right) =~ 2 \,V_p\, T_p ~H. \ee There is an additional conserved
quantity, the reduced angular momentum $J$, which can be obtained in
the following way. Noticing the underlying dynamics to be planar in
the sense $\label{planar} {\dot{\vec{\rho}}}^2 = \dot{\rho}^2 +
\rho^2 \dot{\theta}^2$, then the Lagrangian (\ref{lag}) can be
re-expressed as \be \textsc{L} = -  2 T_p V_p \sqrt{1 -
\left({\dot{\rho}}^2 + \rho^2\dot{\theta}^2\right) } +
\frac{V_p}{2^{7 - p}} \frac{\Omega}{\rho^{7 - p}},\ee which implies
the conserved angular momentum as
 \be \textsc{J} ~=~{\partial
\textsc{L} \over
\partial \dot{\theta}} ~=~ 2 \,V_p \,T_p \,\frac{\rho^2\,
\dot{\theta}}{\sqrt{1-(\dot{\rho}^2+ \rho^2 \,\dot{\theta}^2)}}
~\equiv~ 2 \,V_p \,T_p \,\,J \,, \ee  with the reduced conserved
angular momentum as \be J ~=~ \frac{\rho^2\,
\dot{\theta}}{\sqrt{1-(\dot{\rho}^2+ \rho^2 \,\dot{\theta}^2)}}
\label{angl2}\,\,.\ee With this, we can further express the reduced
Hamiltonian \be \label{effhamiltonian} H = \frac{\sqrt{1 +
\frac{J^2}{\rho^2}}}{\sqrt{1 - \dot{\rho}^2}} -
\frac{\lambda}{\rho^{7 - p}},\ee and the corresponding effective
potential is \be \label{effpotential} V_{\rm eff} = \sqrt{1 +
\frac{J^2}{\rho^2}} - \frac{\lambda}{\rho^{7 - p}},\ee
 which is our
basis for analyzing the dynamical behavior of the system under
consideration in the following section. To solve an orbit explicitly
for allowed  $H$ and $J$, we need the following equation \be
\label{orbiteq}\frac{d \rho}{d\theta} = \pm \frac{\rho^2}{J}
\sqrt{\left(H + \frac{\lambda}{\rho^{7 -p}}\right)^2 - \left(1 +
\frac{J^2}{\rho^2}\right)},\ee where the $\pm$ correlate to the
signs of $\dot{\rho}$ if we assume $J > 0$. The explicit analytical
orbits can be found only for $p = 6$ case and they will be given in
section 4. As we will see, their behaviors are completely consistent
with those obtained in section 3 based on the effective potential
(\ref{effpotential}).

\section{The general dynamical behavior}

Given the preparation of the previous section, we are now ready to
analyze the dynamical behavior of the system under consideration in
the region of interest mentioned earlier. Our starting point is the
effective potential (\ref{effpotential}), \be V_{\rm eff} (\rho) =
\sqrt{1 + \frac{J^2}{\rho^2}} - \frac{\lambda}{\rho^{7 - p}}.\ee To
avoid the brane annihilation from happening, we need to have $ \rho
\gg l_s = \sqrt{\alpha'} = \left(a_p \lambda\right)^{1/(7 - p)}$
where we have used the expression for $\lambda$ given earlier and
$\label{ap} a_p = 4 /[g_s \,\pi^{\frac{5 - p}{2}} \Gamma (\frac{7 -
p}{2})]  \gg 1$ with $g_s \ll 1$ the weak string coupling limit. So
the second term in the effective potential appears to be small but
as we will see it still has important dynamical effect in the region
of interest.

To have some idea on the effective potential behavior\footnote{The
non-linearity has an important consequence in the case of $p = 5$
for which the effective potential develops a maximum when $J^2
> 2 \lambda$ while there is no extremum if it is ignored.
For the rest of cases, this effect doesn't change the characteristic
behavior of the potential whether it is considered or not.}, let us
first ignore the constraint on the allowed range for $\rho$. It is
easy to examine that $V_{\rm eff} \rightarrow 1$ as $\rho
\rightarrow \infty$ for all $p$ under consideration while $V_{\rm
eff} \rightarrow - \infty$ as $\rho \rightarrow 0$ for $p < 6$. For
$p = 6$, $V_{\rm eff} \rightarrow - \infty$ if $J < \lambda$ and
$V_{\rm eff} \rightarrow \infty$ for $J
> \lambda$ as $\rho \rightarrow 0$. As will be seen, only $J >
\lambda$ will be relevant to our interest for this case. So the
above behavior of the potential seems to indicate that for $p < 6$,
its extremum if exists at all should be a maximum, therefore only
unbound orbits exist in the region of interest, while for $p = 6$,
the extremum should be a local minimum, therefore both bound and
unbound orbits exist, depending on the initial energy.

Let us now take a close look at this. The extremum of the potential
can be determined from the vanishing of its first derivative with
respective to $\rho$, i.e., \be \label{extremum}  V'_{\rm eff}
\equiv \frac{d V_{\rm eff}}{d \rho} = - \frac{J^2}{\rho^3}
\left[\frac{1}{\sqrt{1 + \frac{J^2}{\rho^2}}} - \frac{7 -
p}{J^2}\frac{\lambda}{\rho^{5 - p}} \right] = 0. \ee The trivial
extremum occurs at $\rho = \infty$ where the second derivative of
the potential vanishes, and our interest is at the finite one
satisfying \be \label{extremumequation}
 \frac{J^2}{\sqrt{1+\frac{J^2}{\rho^2}}}~=~
(7-p)\frac{\lambda}{\rho^{5-p}}.\label{pte}\ee Assume that the
extremum satisfying the above equation occurs at $\rho = \rho_0$,
let us determine whether the extremum is a local maximum or a local
minimum through the sign of the following second derivative of the
potential at $\rho = \rho_0$: \be V_{\rm eff}'' \equiv
\left.\frac{d^2 V_{\rm eff}}{d \rho^2}\right|_{\rho = \rho_0} = -
\frac{J^2}{\rho_0^3} \left[\frac{\frac{J^2}{\rho_0^3}}{\left(1 +
\frac{J^2}{\rho_0^2}\right)^{3/2}} + \frac{(7 - p)(5 - p)}{J^2}
\frac{\lambda}{\rho_0^{6 - p}}\right].\ee It is obvious that for $p
< 6$, the second derivative is always less than zero at $\rho_0$,
therefore the extremum is a maximum. For $ p = 6$, the extremum
exists at a non-zero finite $\rho_0$ only if\footnote{We actually
need $J \gg \lambda$ to avoid the annihilation. For $J \le \lambda$,
the potential has no extremum at a finite $\rho_0$ and falls either
to zero or to $- \infty$ at small $\rho$, therefore the brane and
anti-brane will inevitably annihilate each other.} $J
> \lambda$ and we can now solve (\ref{extremumequation}) to give \be
\label{p6mininmum} \rho_0 = J \sqrt{\left(\frac{J}{\lambda}\right)^2
- 1}. \ee With this, it is easy to check that the second derivative
is actually positive, therefore giving the extremum a minimum. So
our analysis here is completely consistent with our expectation from
the asymptotical behavior of the potential given above. The
characteristic behaviors of the potential for $p < 6$ and $p = 6$
are given in Fig. 1, respectively.

\begin{figure}
 \psfrag{A}{$V_{\rm eff} (\rho)$}
 \psfrag{B}{$\rho$}
 \psfrag{D}{$\rho_0$}
 \psfrag{C}{$1$}
 \psfrag{E}{($0 \le p < 6$)}
 \psfrag{F}{($p = 6$)}
\begin{center}
\includegraphics{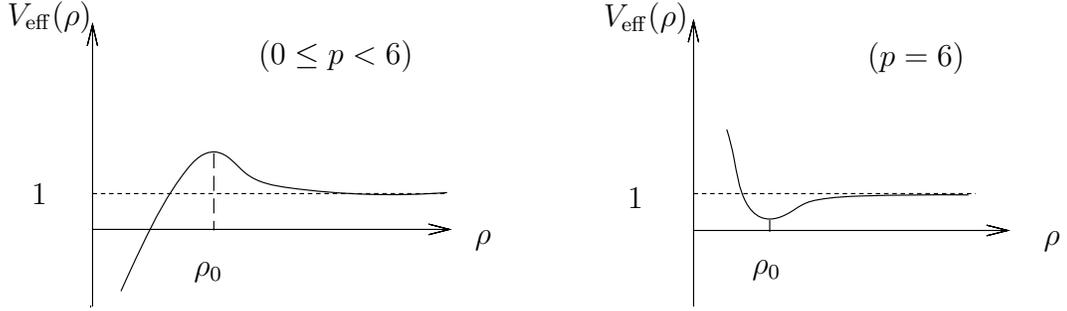}
\end{center}
\caption{The respective characteristic behavior of the potential for
$0 \le p < 6$ (left) and for $p = 6$ when $J > \lambda$ (right).
\label{pb1}}
\end{figure}

Given the above behavior of the potential, we come to analyze the
dynamics of the brane/anti-brane system under consideration in the
region of interest case by case. Let us begin with $p = 6$.

 i) For $p=6$, the extremum occurs at $\rho_0$ as given in
 (\ref{p6mininmum}) when $J > \lambda$
 and the extremum is a minimum. The properties of bound and unbound orbits are shown in Fig.
 2.

 In order to have a bound state, the
 reduced Hamiltonian should satisfy $V_{\rm eff} (\rho_0) \le H < 1$. To avoid the brane annihilation, we need the minimal separation
 (denoted as $\rho^b_{\rm min}$) to satisfy $ \rho^b_{\rm min} \gg l_s = a_6 \lambda$ with $a_6 = 4/g_s \gg 1$. This gives  $ \rho_0 \ge \rho^b_{\rm min} \gg l_s = a_6 \lambda$
which from
 (\ref{p6mininmum}) implies \be \label{p6condition1} J \gg \lambda\, \sqrt{a_6 + \frac{1}{2}} \approx \lambda \,
 \sqrt{a_6}\left(1 + \frac{1}{4 a_6}\right).\ee
 Note that for a bound state ( $\dot{\rho}^b_{\rm min} = 0$)
 \be \label{p6energycon}  H = V_{\rm eff} (\rho^b_{\rm min})
 = \sqrt{1 + \left(\frac{J}{\rho^b_{\rm min}}\right)^2}
 - \frac{\lambda}{\rho^b_{\rm min}} < 1,\ee  which implies
 $\label{p6rhocon1} \rho^b_{\rm min} > \frac{J^2 - \lambda^2}{2
 \lambda}$. So we should have now
 $\rho^b_{\rm min}  > \frac{J^2 - \lambda^2}{2
 \lambda} \gg a_6 \lambda$. This gives \be \label{p6jcon} J \gg
 \lambda\,
 \sqrt{2 \left(a_6 + \frac{1}{2}\right)} \approx \lambda\, \sqrt{2 a_6} \left(1 + \frac{1}{4 a_6}\right), \ee which is a bit stronger than the
 condition (\ref{p6condition1}) as anticipated. Using
 (\ref{p6mininmum}), we have the minimal value of the  effective potential
 as
 \be V_{\rm eff} (\rho_0) = \sqrt{1 -
 \left(\frac{\lambda}{J}\right)^2} \approx 1 - \frac{1}{2} \left(\frac{\lambda}{J}\right)^2,
 \ee which is very close to unit.  So this gives the allowed
 range for the reduced Hamiltonian for bound orbits as
 \be \label{p6energyrange} \sqrt{1 -
 \left(\frac{\lambda}{J}\right)^2} \leq H < 1.\ee So (\ref{p6jcon}) and
(\ref{p6energyrange}) are the conditions  for having  bound orbits
and avoiding the brane annihilation.

\begin{figure}
 \psfrag{A}{$V_{\rm eff} (\rho)$}
 \psfrag{B}{$\rho$}
 \psfrag{E}{$\rho_0$}
 \psfrag{D}{$1$}
 \psfrag{C}{$V_{\rm eff} (\rho_0)$}
 \psfrag{F}{$\rho_{\rm min}$}
 \psfrag{G}{$\rho^b_{\rm min}$}
 \psfrag{H}{$\rho^a_{\rm min}$}
 \psfrag{K}{($p = 5$)}
 \psfrag{L}{($p = 6$)}
\begin{center}
\includegraphics{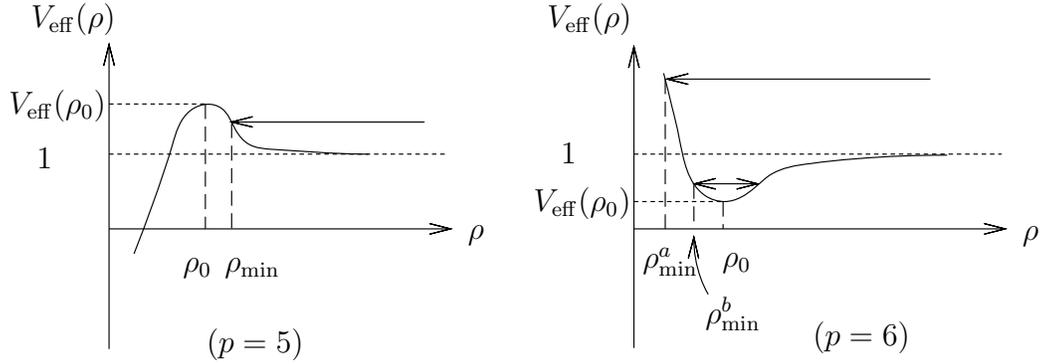}
\end{center}
\caption{The properties of unbound orbits for $p = 5$ when $J^2
> 2 \lambda$ (left) and those of bound and unbound orbits for $p = 6$
when $J \gg \lambda$ (right). For $p = 6$, $\rho^b_{\rm min}$ is the
minimal separation for the bound orbit and $\rho^a_{\rm min}$ is the
correspondence for the unbound orbit. \label{p5p6orbit}}
\end{figure}

We now come to the unbound orbits. Obviously we need to have $H \ge
1$ and this gives \be H = V_{\rm eff} (\rho^a_{\rm min}) = \sqrt{1 +
\left(\frac{J}{\rho^a_{\rm min}}\right)^2}
 - \frac{\lambda}{\rho^a_{\rm min}} \ge 1, \ee where $\dot{\rho}^a_{\rm min} = 0$
 with $\rho^a_{\rm min}$ the present minimal brane separation. The above in turn
implies
 \be \label{p6rhomin} \rho^a_{\rm min} = \frac{J^2 - \lambda^2}
 {\lambda H + \sqrt{J^2 (H^2 - 1) + \lambda^2}} \,\leq \,\frac{J^2 - \lambda^2}{2 \lambda}.\ee
To avoid the brane annihilation, we need also $\rho^a_{\rm min} \gg
l_s = a_6 \lambda$. This, along with (\ref{p6rhomin}), first implies
once again (\ref{p6jcon}) between $J$ and $\lambda$ and then gives a
constraint on $H$, when combined with $H \ge 1$, as \be
\label{p6unboundHcon} 1 \le\, H \,\ll \sqrt{1 + \frac{J^2}{\lambda^2
a_6^2}} - \frac{1}{a_6}.\ee So the conditions for having unbound
orbits and for avoiding the brane annihilation are (\ref{p6jcon})
and (\ref{p6unboundHcon}) for the present case.

ii) For $p=5$, the potential has a possible maximum.
Eq.(\ref{extremum}) implies that the maximum exists if $J^2
> 2 \lambda$ and if so it occurs at \be \label{p5rho0} \rho_0 = \frac{\frac{2
\lambda}{J}} {\sqrt{1 - \left(\frac{2 \lambda}{J^2}\right)^2}}.\ee
The orbits are all unbound and to avoid the brane annihilation, we
need to have  $(a_5 \lambda)^{1/2} \ll \rho_0 < \rho_{\rm min}$ with
$a_5 = 4/g_s \gg 1$. This gives rise to, using the expression for
$\rho_0$ from (\ref{p5rho0}), \be \frac{2}{a_5} \gg
\frac{J^2}{2\lambda}\left[1 -
\left(\frac{2\lambda}{J^2}\right)^2\right].\ee Given  $J^2 > 2
\lambda$ and $a_5 \gg 1$, the above implies that $J^2/(2\lambda)$ is
very close to unit and if we set \be \label{p5jcon1} \frac{J^2}{2
\lambda} = 1 + \epsilon , \ee  then the very small $\epsilon$
satisfies, to leading order,  \be \label{p5jcon2} 0 < \epsilon \ll
\frac{1}{a_5}.\ee In addition, we need to have \be
\label{p5energycon} V_{\rm eff} (\rho_0) > H \ge 1, \ee with now \be
V_{\rm eff} (\rho_0) = \frac{1}{2} \left(\frac{J^2}{2\lambda} +
\frac{2\lambda}{J^2}\right).\ee So the conditions for having unbound
orbits and no occurrence of annihilation for the present case are
(\ref{p5jcon1}), (\ref{p5jcon2}) and (\ref{p5energycon}). The
characteristic features of unbound orbits are given in Fig. 2.

iii) For $p=4$, the potential has a maximum. We have from
(\ref{extremumequation}) the maximum occurring at \be \label{p4rho0}
\rho_0^2 = \frac{2 J^2}{\sqrt{1 + 4 \left(\frac{J^3}{3
\lambda}\right)^2} - 1}.\ee The characteristic features of the
potential and the unbound orbits are similar to those for $p = 5$ as
shown in Fig. 2. We have now \be \label{p4maxpotential} V_{\rm eff}
(\rho_0) = \frac{2^{1/2}\left[1 + \sqrt{1 + 4\left(\frac{J^3}{3
\lambda}\right)^2} + \frac{4}{3}
\left(\frac{J^3}{3\lambda}\right)^2\right]}{\left(1 + \sqrt{1 +
4\left(\frac{J^3}{3 \lambda}\right)^2}\right)^{3/2}}.\ee To have the
unbound orbits and to avoid the brane annihilation, we need both \be
\label{p4energycon} V_{\rm eff} (\rho_0)
> H \ge 1 \ee with $V_{\rm eff} (\rho_0)$ as given above, and the minimal brane separation satisfying $
 \rho_{\rm min} > \rho_0 \gg l_s = (a_4 \lambda)^{1/3}$ with $a_4
= 8/(g_s  \pi ) \gg 1$. This last constraint, combining with the
expression for $\rho_0$ given in (\ref{p4rho0}), gives \be
\label{p4jcon1} 6\sqrt{2}\, \frac{J^3}{3 \lambda} \gg a_4
\left[\sqrt{1 + 4 \left(\frac{J^3}{3\lambda}\right)^2} -
1\right]^{3/2},\ee which implies $J^3/(3\lambda) \ll 1$. To leading
order, we have \be \label{p4jcon2} \frac{J^3}{\lambda} \ll 3
\sqrt{\frac{3}{a_4}} \ll 1,\ee and  the maximum of the potential
from (\ref{p4maxpotential}) can now be approximated as $ V_{\rm eff}
(\rho_0) \approx 1 + \frac{1}{6}
\left(\frac{J^3}{3\lambda}\right)^2$. So the conditions in this case
for having unbound orbits and for avoiding the brane annihilation
are (\ref{p4energycon}) and (\ref{p4jcon2}).

iv) For $ 0 \le p\leq 3$, the situation is similar to $p=4$ case and
to the ($J^2 > 2\lambda$) $p = 5$ case. In other words, the
potential has a maximum. The characteristic features of the
potential and the unbound orbits are similar to those given in Fig.
2. However, the $\rho_0$ for which the potential takes maximum
cannot for now be solved exactly. As will be seen, in the region of
interest, it can still be solved approximately. As usual, to have
unbound orbits and to avoid the brane annihilation, we need both $
V_{\rm eff} (\rho_0)
> H \geq 1$ and $ \rho_{\rm min} > \rho_0
\gg l_s = (a_p \lambda)^{1/(7 - p)}$ with $a_p $ given earlier. Here
$\rho_{\rm min}$ is once again the minimal brane separation for a
given unbound orbit at which $\dot{\rho}_{\rm min} = 0$ and $H =
V_{\rm eff} (\rho_{\rm min})$. The last constraint can be
re-expressed as \be \label{ppcon1} \frac{\lambda}{\rho_0^{7 - p}}
\ll \frac{1}{a_p} \ll 1,\ee and the equation
(\ref{extremumequation}) for determining $\rho_0$ can also be
re-expressed as \be \label{pprho0}
\frac{\frac{J^2}{\rho_0^2}}{\sqrt{1 + \frac{J^2}{\rho_0^2}}} = (7 -
p) \frac{\lambda}{\rho_0^{7 - p}} \ll 1,\ee where we have used the
condition (\ref{ppcon1}). This must imply  $\label{pp1}
\frac{J}{\rho_0} \ll 1$ and therefore to leading order we have from
(\ref{pprho0}) $ \rho_0 \approx J \left[(7 - p) \frac{\lambda}{J^{7
- p}}\right]^{\frac{1}{5 - p}}$. Combining this with the constraint
$\rho_0 \gg (a_p \lambda)^{1/(7 - p)}$, we have \be \label{ppjcon}
\frac{J}{ \lambda^{\frac{1}{7 - p}}} \ll \sqrt{\frac{7 -
p}{a_p^{\frac{5 - p}{7 - p}}}} \ll 1.\ee The maximal value of the
potential can  now be estimated as \be V_{\rm eff} (\rho_0) \approx
1 + \frac{5 - p}{2 (7 - p)^{\frac{7 - p}{5 - p}}} \left(\frac{J^{7 -
p}}{\lambda}\right)^{\frac{2}{5 - p}}
> 1.\ee Note that $p = 4$ is just a special case here but neither $p
= 5$ nor $p = 6$. So the conditions for having unbound orbits and
avoiding the brane annihilation are $V_{\rm eff} (\rho_0) > H \ge 1$
and (\ref{ppjcon}).

In summary, we have analyzed the dynamical behavior of Dp/$\bar {\rm
D}$p system for $0 \le p \le 6$ via its corresponding effective
potential in the region of interest stated earlier. We have
determined the conditions for which the underlying requirements are
met in each case. We find that only for $p = 6$ case, there exist
both bound orbits and unbound orbits and for all other cases
considered only unbound orbits exist. In particular, we find that
the dimensionless ratio $J/\lambda^{1/(7 -p)}$ for $p = 6$, $p = 5$
and the rest cases considered is very different.  Explicitly,
 $J /\lambda \gg 1$ for $p = 6$, $J/\lambda^{1/2}$ is bigger than but
very close to $\sqrt{2}$ for $p = 5$, while $J /\lambda^{1/(7 - p)}
\ll 1$ for $0 \le p \le 4$.  Each of these indicates that the
potential well depth (p = 6) or potential barrier height ( $0 \le p
\le 5$) is very close to its asymptotic value of unit  at $\rho
\rightarrow \infty$, and therefore the stability of the respective
classical orbits needs to be addressed. Given our understanding of
the potential behavior and without going into detail, one expects
that the classically allowed unbound orbits for $0 \le p \le 5$ will
not be stable quantum mechanically as well as non-perturbatively
(for example through tunneling) and even under classical
perturbation, and the brane and anti-brane annihilation seems to be
inevitable. However, for $p = 6$ case, even though the classical
bound orbits and those unbound orbits with $H$ near by unit can
exchange their role under classical perturbation as well as quantum
mechanically, the brane and anti-brane annihilation seems to be
remote\footnote{Our focus here is on the classical orbits and the
other issues of stability such as their possible decays at quantum
level discussed in \cite{Burgess:2003qv} are beyond the scope of the
paper. Even so, given what has been considered there, we expect that
the quantum stability can still hold in the present case since in
our case $g_s N \ll 1$.}. In other words, we don't expect to have
the brane annihilation in this case in general if the reduced
angular momentum $J \gg \lambda$ with $\lambda$ the stringy
parameter.

\section{The ${\rm D}_6/\bar{\rm D}_6$ System}

It is not easy to solve the orbital equation (\ref{orbiteq}) to give
explicit and analytical orbits for a general $p$ ($ 0 \le p \le 6$).
However, this can be done with easy  for $p = 6$. We will  solve the
orbital equation (\ref{orbiteq}) directly for $p = 6$ as an explicit
example, analyze the corresponding dynamical behavior of the orbits
and find the relevant constraints which validate the region of
interest under consideration. We will find that these constraints
are in complete agreement with what we found in the previous section
based purely on the effective potential. For this, let us begin with
(\ref{orbiteq}) for $p = 6$ which is \be \label{p6orbiteq}\frac{d
\rho}{d\theta} = \pm \frac{\rho^2}{J} \sqrt{\left(H +
\frac{\lambda}{\rho}\right)^2 - \left(1 +
\frac{J^2}{\rho^2}\right)},\ee where the $\pm$ correlate to the
signs of $\dot{\rho}$ once we choose $J > 0$, i.e., $\dot{\theta} >
0$. The above equation can be solved to give \be \label{p6sol} \rho
= \left\{\begin{array}{cc} \frac{2\, A}{ Q
\sin[\pm\,\frac{\sqrt{-A}}{J}(\theta-\,\theta_0)]-B}, &\qquad\qquad
J>\lambda\\
\frac{4\,J^2 B}{B^2 (\theta-\,\theta_0)^2 - \,4 J^2 C},&\qquad
\qquad J=\lambda \\  \frac{4 A J \exp[\mp\frac{\sqrt{A}}{J}
(\theta-\,\theta_0) ]}{\left[J
\exp[\mp\frac{\sqrt{A}}{J} (\theta-\,\theta_0) ]- B \right]^2-\,\,4A C}, &\qquad\qquad  J<\lambda\end{array}\right.\\
\ee  where  $\lambda = g_s \sqrt{\alpha'}/4  = \sqrt{\alpha'} / a_6$
and \be\label{ABCQ} A =\lambda^2-J^2,\qquad B = 2\lambda H,\qquad C
= H^2-1,\qquad Q = 2\sqrt{J^2(H^2-1)+\lambda^2}. \ee The reduced
angular momentum and Hamiltonian are given by (\ref{angl2}) and
(\ref{effhamiltonian}) for $p = 6$, respectively. For convenience,
we collect them  here \be J = \frac{\rho^2\,
\dot{\theta}}{\sqrt{1-(\dot{\rho}^2+ \rho^2 \,\dot{\theta}^2)}},
\label{JH} \qquad
 H = \frac{\sqrt{1 +
\frac{J^2}{\rho^2}}}{\sqrt{1 - \dot{\rho}^2}} ~ -
\frac{\lambda}{\rho},\ee with $H > 0$ in the region of interest.

It is not difficult to examine the solutions (\ref{p6sol}) that the
$\rho$ can be made as small as one wishes for large enough $\theta$
for both $J = \lambda$ case and $J < \lambda$ case.  This is also
true for either sign in the latter case. Therefore these orbits
inevitably lead to the annihilation of brane and anti-brane and are
not in the interest of this paper. For this reason, we will then
drop them from further consideration and focus on the case of $J >
\lambda$ from now on.

So the relevant solution is now

\be  \label{p6plussol}\rho  = \frac{- 2\, A}{B-\, Q
\sin[\frac{\sqrt{-A}}{J}(\theta-\,\theta_0)]}\,, \ee\\ where we for
certainty take the `$+$' sign for the $J > \lambda$ case in
(\ref{p6sol}) and the `$-$' sign can be obtained by sending $\theta
- \theta_0 \rightarrow - (\theta -\theta_0)$. Let us analyze the
solution in two separate cases.

~~~~i)~If \,$B>Q$, we have then the following constraint for $H$ as
\be \sqrt{1-\frac{\lambda^2}{J^2}} ~ \leq~ H < ~1 \label{p6Hcon} \ee
where the left side bound is from the non-negative requirement of
the quantity in the square-root of $Q$ in (\ref{ABCQ}) and the
right side bound is from $B > Q$ and the fact $H
> 0$. This is exactly the same constraint
as (\ref{p6energyrange}) derived in the previous section.  It is now
obvious from (\ref{p6plussol}) that $\rho$ has both a maximum and a
minimum as  \bea \rho_{\rm{max}}&=&\frac{-2\,A}{B-\, Q}\,,\qquad
\qquad\qquad {\rm{when}}~~ \theta-\,\theta_0 = \frac{J}{\sqrt{-A}}\,
\frac{\pi}{2} \\\label{rmax} \rho_{\rm{min}}&=&\frac{-2\,A}{B+\,
Q}\,,\qquad \qquad\qquad {\rm{when}}~~ \theta-\,\theta_0 =
\frac{J}{\sqrt{-A}}\, \frac{3 \pi}{2}\label{rmin}. \eea  Note that
the orbit is a bound one but it will not be closed in general unless
$\frac{J}{\sqrt{-A}}= \frac{m}{n}$ with $m,n$ integers, i.e., with a
rational ratio.  With (\ref{ABCQ}) and (\ref{p6Hcon}), we have from
(\ref{rmin})
 \be
\rho_{\rm{min}}=\frac{J^2-\lambda^2}{\lambda H +
\sqrt{\lambda^2-\,J^2(1-\,H^2)}} > \frac{J^2-\lambda^2}{2\lambda}.
\ee  To avoid the brane annihilation, we need the minimal brane
separation to satisfy $ \rho_{\rm{min}}\gg \sqrt{\alpha \,'} = a_6
\lambda $ where in the last equality we have used the expression for
$\lambda$ given earlier. This can be automatically satisfied if we
have $\frac{J^2 -\lambda^2}{2 \lambda} \gg a_6 \lambda$, considering
the above inequality. The resulting constraint is just
(\ref{p6jcon}) derived in the previous section.

~~~~ii)~If \,$0< B \leq Q$, we have $H \ge 1$  from the explicit
expressions of $B$ and $Q$ given in (\ref{ABCQ}) and from the fact
$H > 0$.  It is obvious from the solution (\ref{p6plussol}) that the
orbit is now unbound. We have
 \bea
&&\rho \rightarrow\infty\,, \quad\qquad\qquad\qquad~ {\rm{when}}~~
\theta-\,\theta_0 = \frac{J}{\sqrt{-A}}\,
\left(\pi-\arcsin{\frac{B}{Q}}\right)
\\ &&\label{rminf}\rho_{\rm{min}} =\frac{-2\,A}{B+\,
Q}\,,\,\qquad \qquad {\rm{when}}~~ \theta-\,\theta_0 =
\frac{J}{\sqrt{-A}}\, \frac{3 \pi}{2}\,. \eea To avoid the brane
annihilation, we need to have \be a_6 \lambda \ll \rho_{\rm{min}} =
\frac{J^2-\lambda^2}{\lambda H + \sqrt{J^2(H^2-1)+\lambda^2}} \le
\frac{J^2 - \lambda^2}{2 \lambda},\ee where we have used the
explicit expression for $\rho_{\rm min}$ from (\ref{rminf}) and $H
\ge 1$ for the second inequality which will give the same constraint
between $J$ and $\lambda$ as (\ref{p6jcon}). The constraint on $H$
can be determined from the first inequality above and combining with
$H \ge 1$ we have \be 1 \le\, H \,\ll \sqrt{1 + \frac{J^2}{\lambda^2
a_6^2}} - \frac{1}{a_6}.\ee Once again, we obtain the same
constraints on $J$ and $H$ as those derived in the previous section.

\section{Conclusion}

We study the dynamical behavior of a pair of Dp-brane and anti
Dp-brane ($0 \leq p \leq 6$) moving parallel to each other in the
region where the brane and anti-brane annihilation will not occur
and the low energy description is valid. We find that the classical
orbits can indeed exist under conditions specified in the previous
sections and are in general unbound except for $p = 6$ where bound
orbits can exist. However, these unbound orbits except for $p = 6$
case are in general unstable quantum mechanically as well as
non-perturbatively and even under classical perturbation, and
therefore the brane and anti-brane annihilation seems to be
inevitable. For $p = 6$, we don't expect that the brane and
anti-brane  annihilate each other even though bound orbits and
unbound orbits with $H$ nearby unit can exchange their role under
classical perturbation as well as quantum mechanically. The
non-linearity of DBI action for D-branes plays an important role in
the case of $p = 5$ while it has only a quantitative effect for all
the other cases under consideration.

\section*{Acknowledgements:}

We thank Rong-Jun Wu for reading the manuscript and pointing out a
few typos in the original version. We acknowledge support by grants
from the Chinese Academy of Sciences, a grant from 973 Program with
grant No: 2007CB815401 and grants from the NSF of China with Grant
No:10588503 and 10535060.

\end{document}